%% file: main.tex
\documentclass[conference,a4paper]{IEEEtran}
\usepackage{cite}
\usepackage{amsmath,amssymb,amsfonts}
\usepackage{algorithmic}
\usepackage{graphicx}
\usepackage{textcomp}
\usepackage{threeparttable}
\usepackage{multirow}
\usepackage{colortbl}
\usepackage{tabularray}
\usepackage{booktabs}
\usepackage{bm}

\newcommand{\blue}[1]{{\color{black}#1}}

\begin{document}

\title{Spectral Signatures for Parametric Fault Detection in Flexible Electronics}

\author{\IEEEauthorblockN{Paula Carolina Lozano Duarte\IEEEauthorrefmark{1},
Sule Ozev\IEEEauthorrefmark{2} and
Mehdi Tahoori\IEEEauthorrefmark{1}}
\IEEEauthorblockA{\IEEEauthorrefmark{1}Dept. of Computer Science, Karlsruhe Institute of Technology, Karlsruhe 76131, Germany}
\IEEEauthorblockA{\IEEEauthorrefmark{2}School of Electrical, Computer and Energy Engineering, Arizona State University, Tempe, AZ 85281, USA}
\thanks{This work has been supported by the European Research Council (ERC) (Grant No. 101052764)}
\thanks{Corresponding author: Paula Carolina Lozano Duarte (e-mail: paula.duarte@kit.edu).}}

\maketitle

\input{section/0_Abstract}

\input{section/1_Introduction}
\input{section/2_Background}
\input{section/3_Methodology}
\input{section/4_Results}
\input{section/5_Conclusion}

\section*{Acknowledgment}

This work has been supported by the European Research Council (ERC) (Grant No. 101052764) and the KIT International Excellence Fellowship.

\bibliographystyle{IEEEtran}
\bibliography{references}

\end{document}

%% file: section/0_Abstract.tex
\begin{abstract}
Flexible electronics (FE) based on indium gallium zinc oxide (IGZO) unipolar thin-film transistor (TFT) technologies enable lightweight, conformable systems for wearable sensing, biomedical monitoring, and human-machine interfaces.
These target applications rely on analog and mixed-signal (AMS) processing blocks that must remain accurate despite four key testing challenges: increased device variability and defect rates from low-cost unipolar fabrication; susceptibility to post-manufacturing failures due to the absence of rigid packaging; parametric faults that degrade analog accuracy without hard digital failures; and the high cost and in-field inaccessibility of specialized Automatic Test Equipment (ATE) for low-cost disposable FE systems. 
Conventional test approaches relying on analog-to-digital converters (ADCs) incur prohibitive area overhead in resource-constrained systems; critically, ADC accuracy is itself subject to process, voltage, and temperature (PVT) variations, leading to test invalidations.
This work proposes a frequency-domain test signature generation circuit exploiting spectral energy distribution from a voltage-controlled oscillator (VCO) to detect faults and parametric deviations in AMS circuits under test (CUTs).
The VCO converts fault-induced control-voltage deviations into frequency shifts processed through lightweight analog filters.
Root-mean-square (RMS) values of harmonic components serve as compact signatures exhibiting monotonic sensitivity to operating-point deviations while preserving ordering across PVT corners, ensuring robust operation without precision references or matched comparators.
\blue{A ratiometric refinement of this signature is shown to suppress post-calibration drift from temperature and supply variation by up to an order of magnitude, validated through direct monotonicity verification across all evaluated corners and supply voltages.}
The signature is readable via two standard I/O pins, enabling in-field test without ATE.
The circuit occupies 6390~$\mu$m$^2$ and consumes 16.7~$\mu$W, achieving 26$\times$ area reduction versus compact ADCs and nearly 1500$\times$ versus SAR implementations.
\end{abstract}
\begin{IEEEkeywords}
Flexible electronics, voltage-controlled oscillator,
parametric fault detection, mixed-signal test
\end{IEEEkeywords}

%% file: section/1_Introduction.tex
\section{Introduction}
\label{sec:intro}

Flexible electronics (FE) based on indium gallium zinc oxide (IGZO) thin-film transistors (TFTs) enable mechanically compliant systems on plastic substrates~\cite{EuropracticeFlexibleElectronics} for wearable health monitoring~\cite{afentaki2025stress}, large-area sensing~\cite{Zhang2025}, electronic skin~\cite{Leogrande2025}, flexible displays~\cite{Lim2025}, and human-machine interfaces~\cite{Heng:AM2022:FlexHumanMachInterfaces}.
These applications rely on analog and mixed-signal (AMS) front-end blocks whose parametric accuracy is critical for correct system operation.

FE systems present four testing challenges absent in conventional silicon VLSI. First, low-cost unipolar fabrication---supporting only n-type devices at larger feature sizes---yields higher device variability and defect rates~\cite{Han2025}.
Second, lack of rigid packaging exposes systems to post-manufacturing mechanical failures during in-field use~\cite{Singh:BIST_FE}.
Third, the prevalence of analog blocks means parametric faults---threshold shifts, mobility degradation, resistive deviations---silently degrade accuracy without hard digital failures~\cite{bilgic2022performance}.
Fourth, the high cost of specialized ATE and its inaccessibility after deployment make it unsuitable for low-cost disposable FE systems and preclude in-field testing.
These factors collectively demand compact, low-power, on-chip test support circuits for unipolar TFT technologies.

Conventional mixed-signal test for analog circuits commonly relies on ADC-based voltage monitoring~\cite{Oshita2016, Jeong2016}.
In unipolar TFT processes, this is problematic: matched comparators are highly sensitive to threshold-voltage spread; stable references require unavailable p-type devices; and pseudo-CMOS digital logic incurs large area and static power~\cite{afentaki2025stress}.
Critically, ADC accuracy degrades under the supply fluctuations characteristic of FE, causing the test circuit itself to produce incorrect pass/fail decisions.
Oscillator-based test schemes exploit delay a frequency sensitivity to detect defects~\cite{lu2025testingfaulttolerancetechniques, LiROtesting}, but time-domain counting requires digital counters and stable reference clocks, which are impractical in low-frequency FE; and absolute frequency is strongly PVT-dependent in IGZO.
Exploiting frequency-domain spectral energy as a lightweight on-chip test observable 
remains unexplored in unipolar TFTs.

This work proposes a frequency-domain test signature generation circuit based on an IGZO VCO, detecting faults in \emph{analog CUTs} through spectral energy redistribution while remaining robust against PVT variations.
The generated signature is readable via two standard I/O pins by a microcontroller already present in wearable systems, enabling in-field test without ATE.
\textbf{The key contributions of this work are:}
\begin{itemize}
  \item A VCO-based frequency-domain test signature circuit for parametric fault detection in analog and AMS CUTs, requiring no ADCs, precision references, or CUT modification.
  \item Definition and validation of ratiometric RMS spectral signatures that maintain fault separability across global PVT variations, ensuring reliable pass/fail decisions under large device variability.
  \item Validated implementation in PragmatIC's FlexIC IGZO process, occupying 6390~$\mu$m$^2$ and consuming 16.7~$\mu$W.
  \blue{\item Experimental verification of signature monotonicity across all PVT corners and supply voltages, and a ratiometric refinement that suppresses post-calibration temperature and supply drift by up to an order of magnitude.}
\end{itemize}

%% file: section/2_Background.tex
\section{Background}
\label{sec:background}

\subsection{Flexible Electronics and IGZO TFTs}

FE encompass systems manufactured on mechanically compliant substrates, enabling operation under bending or twisting without functional degradation~\cite{EuropracticeFlexibleElectronics, Han2025}.
FE processes prioritize mechanical robustness and low-temperature compatibility over aggressive scaling, adapting deposition and patterning to substrates such as polyimide.

Among FE materials, IGZO TFTs offer relatively high carrier mobility, optical transparency, and low-temperature processing compatibility~\cite{Zhu:IGZO2021}.
However, IGZO inherently supports only n-type devices, imposing fundamental circuit design constraints, necessitating pseudo-CMOS implementations with increased static power~\cite{high-speed_yuanfeng, lozano2026AKANCOdesig}.
Furthermore, IGZO TFTs exhibit pronounced device variability---originating from process non-uniformities in carrier mobility, threshold voltage, and contact resistance---that directly impacts analog operating points and circuit reliability~\cite{Pan:IGZOTFT2024, pal2026VTS, lozano2026ETS}.
Post-manufacturing failures due to mechanical stress further increase the effective defect rate during in-field operation~\cite{Singh:BIST_FE}.
These characteristics make IGZO well-suited for low-power wearable sensing, while simultaneously motivating dedicated test strategies to manage its inherent variability.




VCOs are a common functional block in FE
sensing platforms, providing tunable timing references for synchronization and control~\cite{circuits_A-IGZO_Mallory, A-igzo_di, fastiIC_arun}.
Existing IGZO TFT VCO implementations face trade-offs among tunability, power, and integration~\cite{VCO_igzo_tejaswini, VCO_Bongjun, lozanoduarte2026lowpowerpllbasedclockstabilization}, yet their intrinsic sensitivity to device parameters and bias conditions, presents an opportunity for test, motivating their use as on-chip test transducers.

\subsection{AMS Test andW Built-In Self-Test Architectures}

Conventional AMS test relies on parametric measurements comparing analog quantities against
limits~\cite{bilgic2022performance}.
This often requires ADCs and digital post-processing, incurring significant development cost and ATE dependence~\cite{Oshita2016}.
BIST techniques embed test functionality on-chip, reducing external access and enabling in-field testing~\cite{Jeong2016}.
Not all analog BIST methods are ADC-based: lightweight structural approaches using oscillators and delay monitors have been proposed~\cite{kashyap2024structural, bilgic2023low}.
Nevertheless, ADC-driven architectures dominate AMS BIST due to their direct voltage observability.

In unipolar TFTs, ADC-based BIST faces compounded challenges:
matched comparators are sensitive to threshold variations severe in IGZO; stable references are difficult without complementary devices. 
Critically, since the BIST circuitry is fabricated in the same variable process as the CUT, ADC accuracy degrades under supply-voltage fluctuations characteristic of FE, causing \emph{test invalidations}---incorrect pass/fail decisions produced by the BIST itself.

Delay- and oscillator-based BIST schemes exploit propagation delay sensitivity to defects~\cite{lu2025testingfaulttolerancetechniques, LiROtesting}.
In FE, \cite{Singh:BIST_FE} presents BIST for interconnect reliability under mechanical stress.
However, these rely on time-domain or absolute frequency measurements requiring digital counters and reference clocks.
Exploiting frequency-domain spectral energy distribution as a lightweight on-chip test signature---readable via standard I/O without ATE or counter complexity---remains unexplored in unipolar TFTs, motivating this work.

%% file: section/3_Methodology.tex
\section{Spectral Test Signature Methodology}
\label{sec:methodology}

\subsection{Rationale and Operating Principle}

\begin{figure}[t]
\centering
\includegraphics[width=0.95\linewidth]{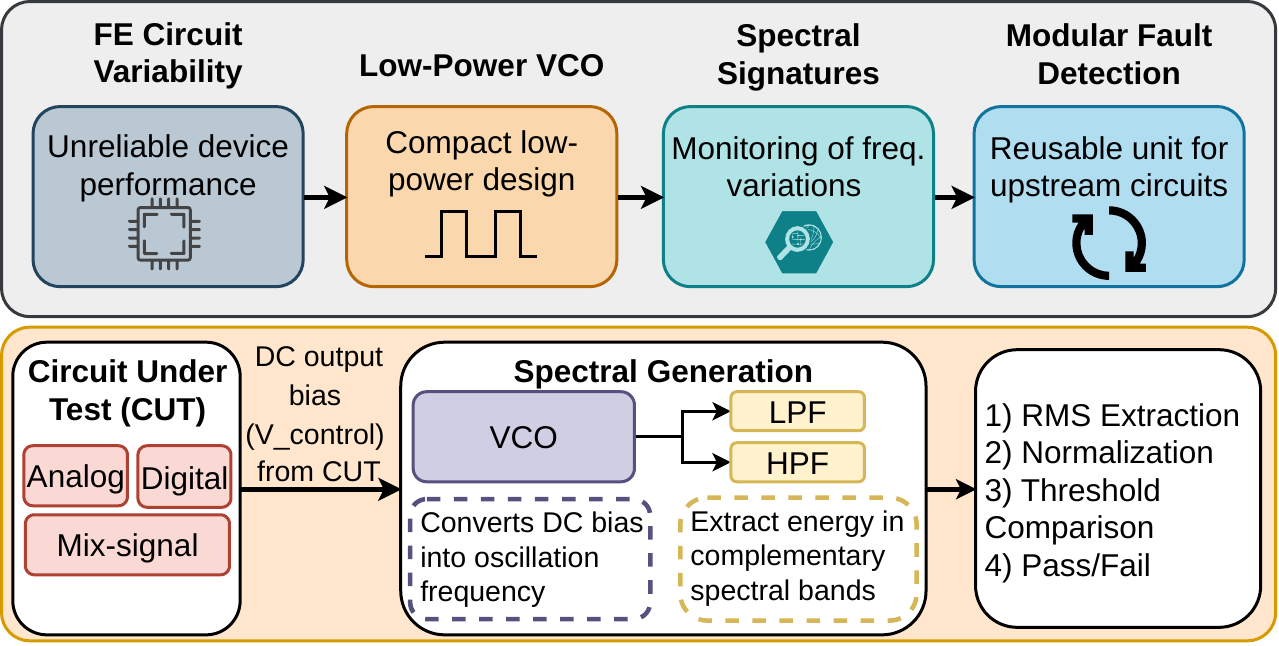}
\caption{Complete signal flow of the proposed frequency-domain test signature circuit. The CUT output bias drives the VCO; LPF and HPF extract spectral energy; normalized RMS signatures are compared against threshold $\tau$ to produce a binary pass/fail decision without ADC or digital counter.}
\label{fig:bist_flow}
\end{figure}

Fig.~\ref{fig:bist_flow} illustrates the complete signal flow of the proposed circuit.
During test, the CUT is set at its nominal input stimulus; its DC output bias---which deviates from the fault-free value in the presence of a parametric fault---drives $V_{\text{CONTROL}}$ of the VCO.
The VCO translates this DC bias into an oscillation frequency; a fault-induced bias shift changes the frequency and redistributes spectral energy across harmonics.
The LPF and HPF extract energy from complementary spectral bands; their RMS outputs are normalized to a stored fault-free reference $S_\text{ref}$ obtained during a one-time calibration step after manufacturing.
If the normalized signature $S_\text{norm}$ exceeds a predefined threshold $\tau$, a FAIL is declared.
The filters output voltages constitute the test signature and can be read via standard I/O pins by any microcontroller present in the system.
\blue{The top row of Fig.~\ref{fig:bist_flow} summarizes the underlying rationale: device variability in the FE circuit motivates a low-power VCO transducer, whose frequency variations are monitored as spectral signatures, enabling modular fault detection that can be reused across multiple upstream analog blocks without modification.}

The methodology requires no frequency measurement, no digital counters, and no voltage digitization.
Only monotonic and repeatable relationships between the control voltage and the extracted signatures are required. 
The design follows three principles: reuse of an existing on-chip VCO as the test transducer; extraction of compact spectral energy features rather than full spectral decomposition; and ratiometric normalization to suppress PVT variability in the test circuitry itself.

\subsection{Test Circuit Architecture and Components}

\subsubsection{VCO as Voltage-to-Frequency Transducer}

The VCO is the sensing core of the proposed circuit. It is implemented using a conventional ring-oscillator topology adapted for unipolar IGZO TFTs (Fig.~\ref{fig:circuit}).
Although the ring-oscillator topology itself is not novel, its low-power operation in the FlexIC process and its reuse as a test transducer rather than solely as a clock source constitute the design contribution of this work.
The VCO converts the DC control voltage $V_\text{CONTROL}$, driven by the CUT output, into an oscillatory signal whose frequency depends on device delays, bias conditions, and environmental parameters.
A fault in the CUT shifts $V_\text{CONTROL}$, thereby altering the oscillation frequency and redistributing the spectral energy of the VCO output.

\subsubsection{Spectral Signature Extraction (LPF and HPF)}

Two first-order RC filters process the VCO output:
\begin{itemize}
  \item \textbf{LPF}: captures near-fundamental spectral content, providing a stable reference that tracks the nominal oscillation regime.
  \item \textbf{HPF}: emphasizes higher-order harmonic components that become more pronounced when oscillation frequency shifts or waveform symmetry is disturbed by a fault.
\end{itemize}

The cutoff frequencies are defined as:
\begin{equation}
  f_\text{LPF} \in [0,\, f_\text{VCO,low}], \quad
  f_\text{HPF} \in [f_\text{VCO,high},\, \infty)
\end{equation}
where $f_\text{VCO,low}$ and $f_\text{VCO,high}$ bracket the nominal VCO operating range.
The RMS values of the filtered signals serve as compact energy-related signatures:
\begin{equation}
  S_\text{LPF} = \text{RMS}(v_\text{LPF}), \quad
  S_\text{HPF} = \text{RMS}(v_\text{HPF})
\end{equation}
\blue{where $v_\text{LPF}$ and $v_\text{HPF}$ denote the time-domain voltage waveforms at the LPF and HPF outputs, respectively.}
\blue{While $S_\text{HPF}$ alone is used for fault detection in Section~\ref{sec:methodology}-3, $S_\text{LPF}$ is additionally used in Section~\ref{sec:results}-E to form the ratio metric $R=S_\text{HPF}/S_\text{LPF}$, which improves robustness to post-calibration environmental drift by exploiting the common-mode sensitivity of both filter outputs to temperature and supply variations.}
Because they integrate spectral energy over a band, RMS metrics are less sensitive to instantaneous noise and waveform jitter than time-domain measurements.

\subsubsection{Pass/Fail Decision and PVT Tolerance}

The pass/fail decision is based on the normalized
signature:
\begin{equation}
  S_\text{norm} = \frac{S_\text{HPF}}{S_\text{ref}}
\end{equation}
where $S_\text{ref}$ is the HPF RMS value measured at the fault-free operating point for each corner during a one-time post-manufacturing calibration step.
\blue{The fault-free operating point is defined, for a given CUT, as the DC output bias the CUT produces under its specified nominal input stimulus and supply conditions, in the absence of any parametric fault; this value is CUT-specific and determined at design time from the CUT's nominal transfer characteristic, not from the BIST circuit itself.}
This per-corner normalization eliminates the absolute PVT spread (up to 5$\times$ between extreme corners), collapsing all corners into a narrow normalized band.
A fault is declared when:
\begin{equation}
  \text{Decision} =
  \begin{cases}
    \text{FAIL} & \text{if } S_\text{norm} > \tau_\text{high}
                  \text{ or } S_\text{norm} < \tau_\text{low}\\
    \text{PASS} & \text{otherwise}
  \end{cases}
\end{equation}
The thresholds $\tau_\text{high}$ and $\tau_\text{low}$ must bracket the worst-case normalized PVT spread (to avoid false positives) while remaining sensitive to the minimum detectable fault (to ensure true positives).
\blue{The RMS computation and threshold comparison are performed off-chip: the LPF and HPF output voltages constitute the test signature and are read directly via two standard I/O pins, with $S_\text{LPF}$, $S_\text{HPF}$, and the pass/fail decision computed by a microcontroller already present in the target system, avoiding on-chip RMS detectors or comparators.}
In the high/medium-sensitivity region ($V_\text{CONTROL} < 1.8$~V), the worst-case PVT spread is $\leq$9.5\%, while a 300~mV fault produces a 22\% signature change, yielding a margin of $>$12\%.
Setting $\tau_\text{high}=1.10$ and $\tau_\text{low}=0.90$ satisfies both conditions across all evaluated PVT corners.
The LPF output provides a complementary low-sensitivity reference, reducing susceptibility to false positives, while the HPF output maximizes sensitivity to fault-induced harmonic distortion.

\subsection{Circuit Implementation}

The complete architecture integrates the VCO and output processing blocks into a compact on-chip structure.
Fig.~\ref{fig:circuit} illustrates the circuit-level implementation.
No external clock, frequency counter, ADC, or high-speed digital logic is required.
The LPF and HPF are realized as first-order RC filters with cutoffs selected to separate fundamental and harmonic content.
The system is fully compatible with the low-frequency, low-power, and large-feature constraints of IGZO-based FE and other unipolar TFT technologies.

%% file: section/4_Results.tex
\section{Simulation Results and Evaluation}
\label{sec:results}

\subsection{Simulation Setup}

All simulations were performed using Cadence Spectre with device models from PragmatIC's third-generation FlexIC process~\cite{EuropracticeFlexibleElectronics}.
The complete test circuit operates from a single supply voltage of $V_{DD} = 3$~V.
The VCO control voltage ($V_{\text{CONTROL}}$) was swept from 1.2~V to 2.7~V to cover the full intended operating range.
\blue{Supply voltage was additionally varied to $V_{DD} \in \{2.7, 3.0, 3.3\}$~V (i.e., $\pm10\%$ around nominal) to evaluate robustness to realistic battery-supply fluctuations.}

PVT variations were evaluated using the available corner models, including typical (T), fast (F), and slow (S) devices at three temperature points (T0, T1, T2).
Results are normalized to the fault-free corner (T1\_T) at $V_{\text{CONTROL}} = 1.2$~V, allowing relative trends to be emphasized over absolute magnitudes.
Device dimensions and resistance values are summarized in Table~\ref{tab:sizing}.

Periodic steady-state (PSS) simulations were used to obtain the steady oscillatory behavior of the VCO.
From the steady-state solution, spectral components were extracted and processed through LPF and HPF filters.
The RMS values of selected harmonic components at the filter outputs were computed and used as compact, energy-related test signatures.

\begin{figure}[t]
    \centering
    \includegraphics[width=0.85\linewidth]{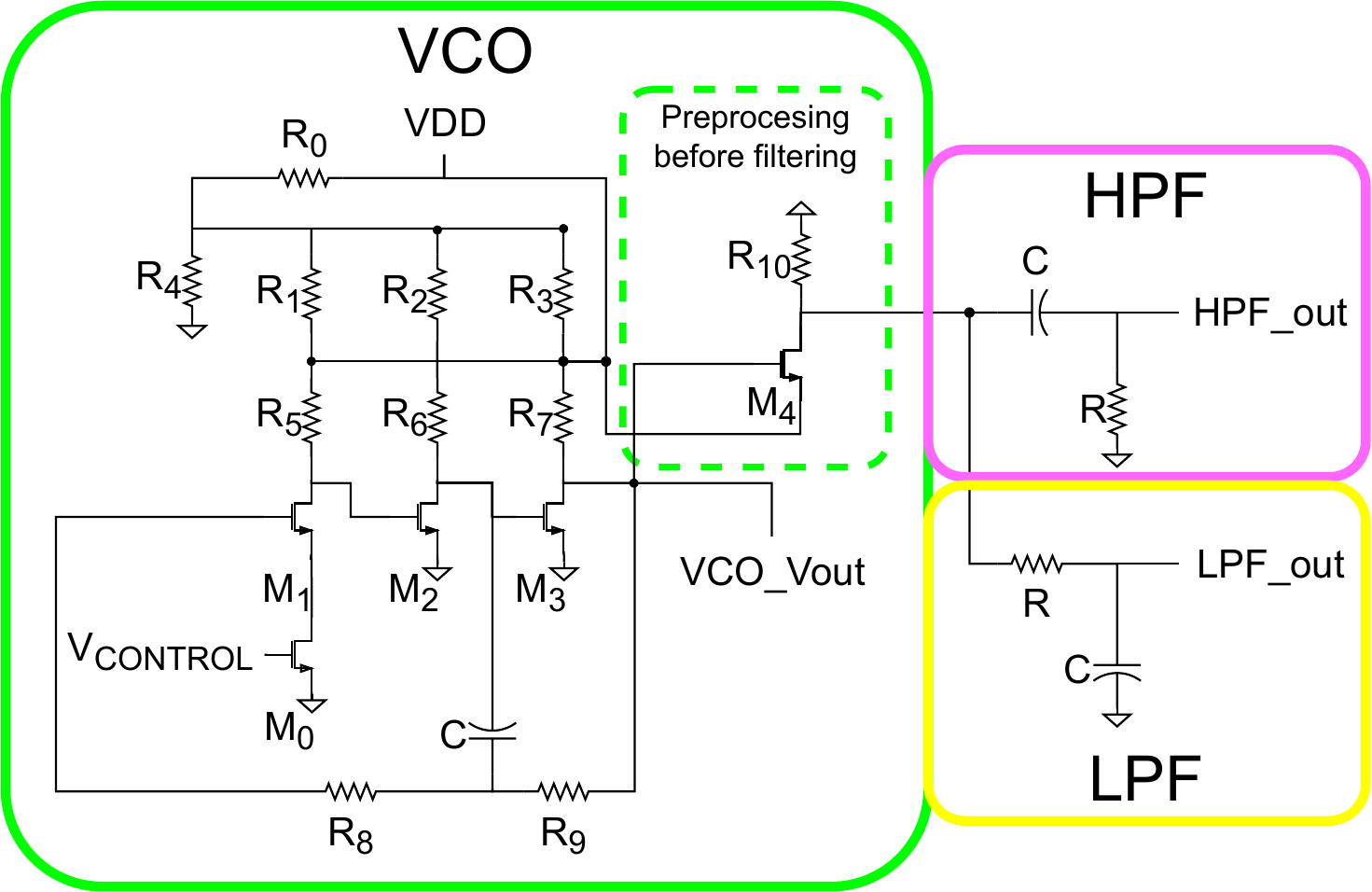}
    \caption{Circuit-level implementation of the proposed frequency-domain test signature circuit, including the IGZO-based VCO and the LPF/HPF output processing stages.}
    \label{fig:circuit}
\end{figure}

\begin{table}[t]
\centering
\caption{Component Sizing Specifications}
\label{tab:sizing}
\scalebox{1}{\input{tables/sizing}}
\end{table}

\subsection{Fault Modeling and Circuit Under Test}

The proposed circuit targets parametric faults in analog CUTs whose output bias voltage drives $V_{\text{CONTROL}}$.
Representative CUTs in FE wearable sensing systems include source-follower bias stages and resistive sensor readout amplifiers, where the output DC level is sensitive to threshold-voltage shifts, mobility degradation, and resistive component deviations common in IGZO processes.
\blue{Two distinct mechanisms justify the evaluated fault range.
First, prolonged positive bias stress in IGZO TFTs induces threshold-voltage shifts of 100--300~mV through charge trapping at the channel/dielectric interface~\cite{Pan:IGZOTFT2024}, a well-characterized degradation mode under sustained operation.
Second, contact or via resistance increases---caused by mechanical strain or electromigration in flexible interconnects---directly add a resistive drop to the source-follower output node, producing deviations in the same range for typical bias currents.
In a source-follower stage, both mechanisms translate directly into an equal DC output deviation, since $\Delta V_\text{out} \approx \Delta V_\text{th}$ in weak-inversion operation.
This places both fault mechanisms within the evaluated range of 75--450~mV.}
In all such cases, a parametric fault manifests as a deviation $\Delta V$ in the DC output voltage driving the VCO control input.

Following established analog fault simulation practice~\cite{spence:analogfaultsimulation}, faults are modeled as controlled additive deviations in $V_{\text{CONTROL}}$ relative to the fault-free operating point $V_0$.
Four fault magnitudes are evaluated: $|\Delta V| \in \{75, 150, 300, 450\}$~mV, corresponding to approximately 5\%, 10\%, 20\%, and 30\% of the full 1.2--2.7~V operating range.
Each magnitude is applied at three representative operating points---$V_0 = 1.2$~V (high-sensitivity), $V_0 = 1.5$~V (intermediate), and $V_0 = 2.0$~V (low-sensitivity)---and evaluated across all nine PVT corners.

\subsubsection{Monotonicity Verification}

\blue{The proposed methodology requires a monotonic and repeatable relationship between $V_\text{CONTROL}$ and the extracted signature; this assumption was verified directly rather than assumed.
Across all nine PVT corners and three supply voltages (27 corner$\times$VDD combinations), $S_\text{HPF}$ was found to be strictly monotonic in $V_\text{CONTROL}$ for all combinations \emph{above} $V_\text{CONTROL}=1.125$~V.
Below this point, a small non-monotonic dip is observed in 8 of the 27 combinations, predominantly in slow- and fast-process corners.
This places the negative-polarity fault evaluation at $V_0=1.2$~V---specifically $|\Delta V| \geq 150$~mV, which probes $V_\text{CONTROL} \leq 1.05$~V---partially within this non-monotonic region.
Table~\ref{tab:fault_coverage_polarity} reports fault coverage separately by polarity at $V_0=1.2$~V to isolate this effect: positive-polarity coverage remains valid and representative, while negative-polarity coverage at $|\Delta V| \geq 150$~mV should be interpreted with this caveat.
Operating points $V_0 \geq 1.5$~V remain entirely within the verified monotonic region for all evaluated fault magnitudes.
}

\begin{figure*}[t]
\centering
\includegraphics[width=0.95\linewidth]{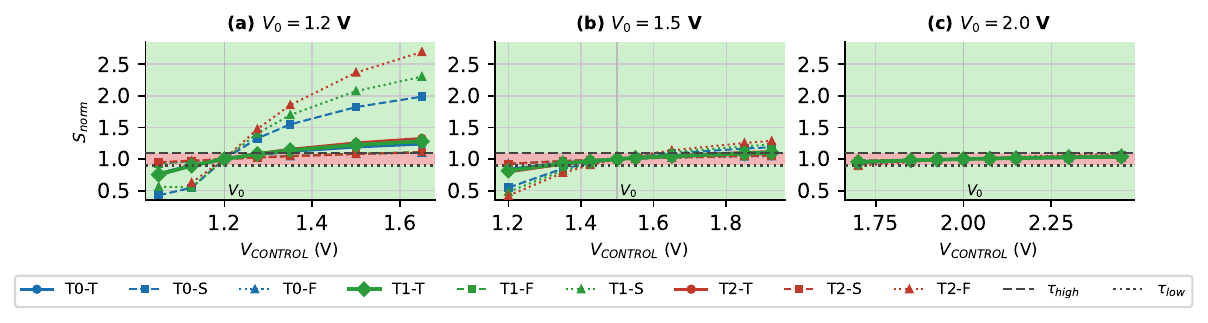}
\caption{Normalized RMS signatures $S_\text{norm}$ under fault injection for all nine PVT corners at three operating points: (a)~high-sensitivity region $V_0=1.2$~V, (b)~intermediate region $V_0=1.5$~V, and (c)~low-sensitivity region $V_0=2.0$~V. Dashed and dotted horizontal lines indicate the pass/fail thresholds $\tau_\text{high}=1.10$ and $\tau_\text{low}=0.90$. The green region represents regions where faults are detected, and the red one is the undetectable zone.}
\label{fig:fault_injection}
\end{figure*}

Table~\ref{tab:fault_coverage} summarizes the Fault Coverage (FC) using a two-sided normalized threshold ($\tau_\text{high}=1.10$, $\tau_\text{low}=0.90$), evaluated across all nine PVT corners and both fault polarities.
Three operating points are compared to illustrate the effect of the sensitivity region on detection capability.
In the high-sensitivity region ($V_0=1.2$~V), FC reaches 100\% for $|\Delta V|=450$~mV and 65\% for $|\Delta V|=150$~mV.
At the intermediate point ($V_0=1.5$~V), FC drops to 76\% and 28\% for the same magnitudes, reflecting the reduced VCO gain in this region.
In the low-sensitivity region ($V_0=2.0$~V), FC remains near zero for all fault magnitudes $\leq 300$~mV.
The incomplete coverage at small fault magnitudes is attributed to fast-process corners (T0\_F, T1\_F, T2\_S), where higher absolute VCO gain reduces relative sensitivity: a 300~mV fault produces only $\approx$8\% change in $S_\text{norm}$ for these corners, below $\tau_\text{high}$.
Slow-process corners (T0\_S, T1\_S, T2\_F) are highly sensitive, with $S_\text{norm}$ reaching 2.37 for a $+300$~mV fault at $V_0=1.2$~V.

\begin{table}[t]
\centering
\caption{Fault Coverage (FC) vs.\ Fault Magnitude ($|\Delta V|$) and Operating Point}
\label{tab:fault_coverage}
\begin{threeparttable}
\resizebox{\linewidth}{!}{
\centering
\setlength{\arrayrulewidth}{0.4pt}
\begin{tblr}{
  colspec = {Q[90] Q[90] Q[90] Q[90]},
  vline{2,3,4} = {2,3,4,5}{0.4pt},
  rowsep = 1.15pt,
  hline{1,2,6} = {1.2pt},
  hline{3,4,5} = {-}{0.4pt},
}
\textbf{$|\Delta V|$} &
\textbf{FC at $1.2$~V} &
\textbf{FC at $1.5$~V} &
\textbf{FC at $2.0$~V} \\
75~mV  (5\%)  & 44\,\% &  0\,\% &  0\,\% \\
150~mV (10\%) & 65\,\% & 28\,\% &  0\,\% \\
300~mV (20\%) & 69\,\% & 56\,\% &  6\,\% \\
450~mV (30\%) & 100\,\% & 76\,\% & 17\,\% \\
\end{tblr}
}
\begin{tablenotes}\footnotesize
\item[] $\tau_\text{high}\!=\!1.10$, $\tau_\text{low}\!=\!0.90$; evaluated across 9 PVT corners $\times$ 2 fault polarities.
\end{tablenotes}
\end{threeparttable}
\end{table}

\begin{table}[t]
\centering
\caption{Fault Coverage at $V_0=1.2$~V, Split by Polarity}
\label{tab:fault_coverage_polarity}
\begin{threeparttable}
\resizebox{\linewidth}{!}{
\centering
\setlength{\arrayrulewidth}{0.4pt}
\begin{tblr}{
  colspec = {Q[80] Q[70] Q[70] Q[120]},
  vline{2,3,4} = {2,3,4,5}{0.4pt},
  rowsep = 1.15pt,
  hline{1,2,6} = {1.2pt},
  hline{3,4,5} = {-}{0.4pt},
}
\textbf{$|\Delta V|$} &
\textbf{FC(+)} &
\textbf{FC($-$)} &
\textbf{Note} \\
75~mV  & 33\,\% & 56\,\% & both monotonic \\
150~mV & 67\,\% & 62\,\% & ($-$) partially affected \\
300~mV & 67\,\% & 75\,\% & ($-$) partially affected \\
450~mV & 100\,\% & N/A & ($-$) non-monotonic region \\
\end{tblr}
}
\begin{tablenotes}\footnotesize
\item[] Results obtained at $V_0=1.2$~V. Negative faults become non-monotonic for $|\Delta V|=450$~mV.
\end{tablenotes}
\end{threeparttable}
\end{table}

\subsection{VCO Frequency Characteristics}

\begin{figure}[t]
\centering
\includegraphics[width=0.85\linewidth]{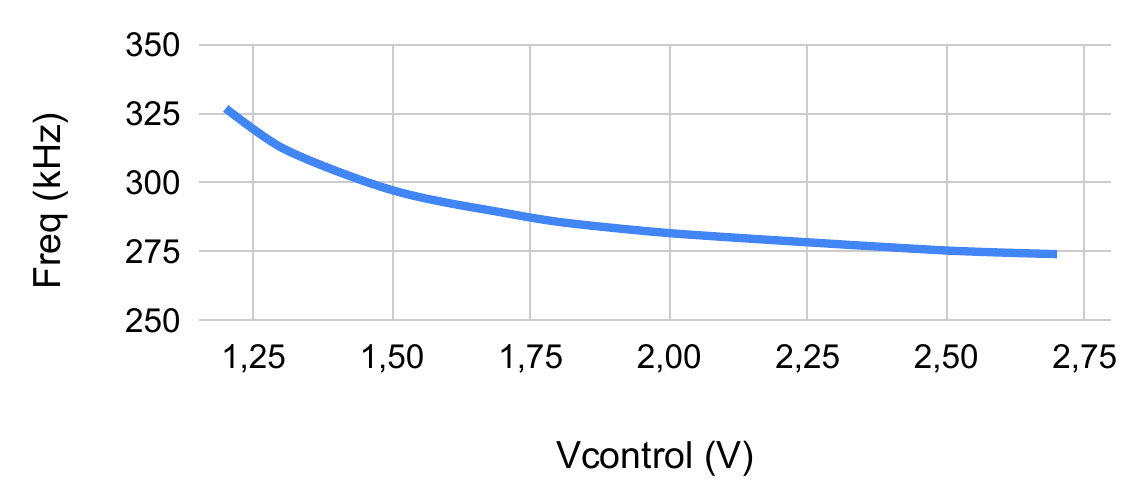}
\caption{Oscillation frequency of the proposed VCO as a function of $V_{\text{CONTROL}}$ under fault-free PVT conditions.}
\label{fig:vco_freq}
\end{figure}

\begin{figure}[t]
\centering
\includegraphics[width=0.95\linewidth]{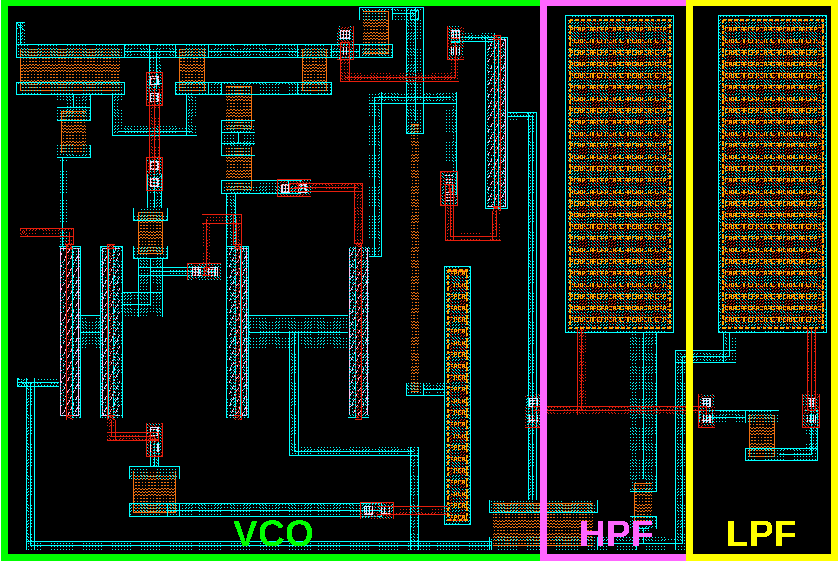}
\caption{Physical layout of the proposed frequency-domain test signature circuit, highlighting the circuit blocks.}
\label{fig:layouts}
\end{figure}

Fig.~\ref{fig:vco_freq} illustrates the oscillation frequency as a function of $V_{\text{CONTROL}}$ under fault-free PVT conditions.
The VCO exhibits a monotonic tuning characteristic across the full control-voltage range, with higher oscillation frequencies at lower control voltages.
This behavior is consistent with the control mechanism described in Section~\ref{sec:methodology}, where $V_{\text{CONTROL}}$ modulates the effective discharge path of the first inverter stage and thus the stage delay.

Although the frequency--voltage relationship is not strictly linear, its monotonic behavior \blue{above $V_\text{CONTROL}=1.125$~V (Section~\ref{sec:results}-B)} ensures an unambiguous correspondence between the control voltage and the oscillation regime, which is sufficient for the proposed methodology within that range.
The physical layout is shown in Fig.~\ref{fig:layouts}, demonstrating a compact implementation occupying 6390~$\mu$m$^2$.
\blue{Comparing standalone VCO operation against the filter-loaded configuration at the same corner and $V_\text{CONTROL}$, the oscillation frequency deviates by less than 0.85\%, confirming that LPF/HPF signature extraction does not measurably degrade the VCO's native transducer behavior.}

\subsection{Frequency-Domain Signatures Under Fault-Free Conditions}

\begin{figure}[t]
\centering
\includegraphics[width=\linewidth]{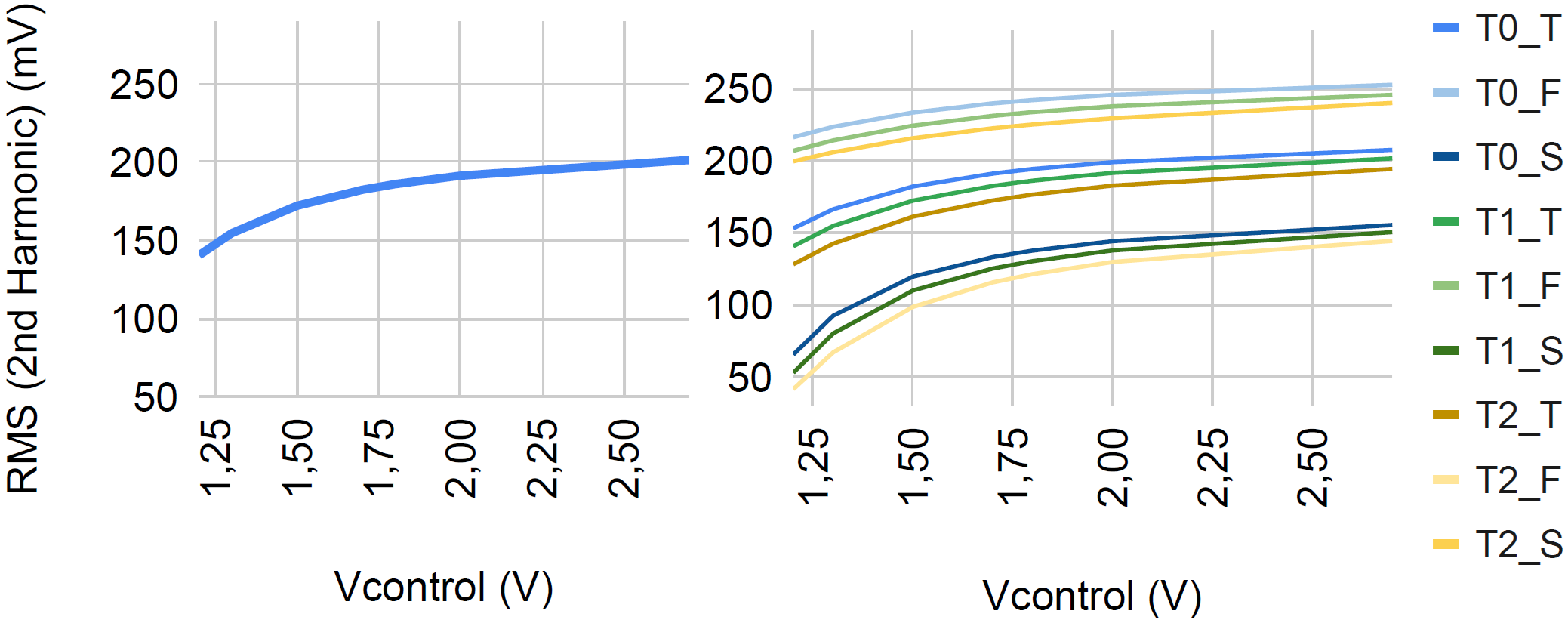}
\caption{RMS value of the second harmonic versus $V_{\text{CONTROL}}$ under nominal (left) and PVT-varying conditions (right).}
\label{fig:rms_combined}
\end{figure}

Fig.~\ref{fig:rms_combined} reports the RMS value of the second harmonic under fault-free PVT conditions.
A clear monotonic increase is observed, with values rising from 0.14 at $V_{\text{CONTROL}}=1.2$~V to 0.2 at $V_{\text{CONTROL}}=2.7$~V.
Sensitivity is non-uniform: in the lower region (1.2--1.5~V) the RMS signature increases by approximately 22.5\% (from 0.14 to 0.17), while in the higher region (2.0--2.7~V) the increase is only 5.2\% (from 0.19 to 0.2).
This confirms that frequency-domain RMS metrics provide a compact, monotonic indicator of the VCO operating point, suitable for detecting deviations induced by faults.

\subsection{Robustness Under PVT Variations}

\begin{figure}[t]
\centering
\includegraphics[width=1\linewidth]{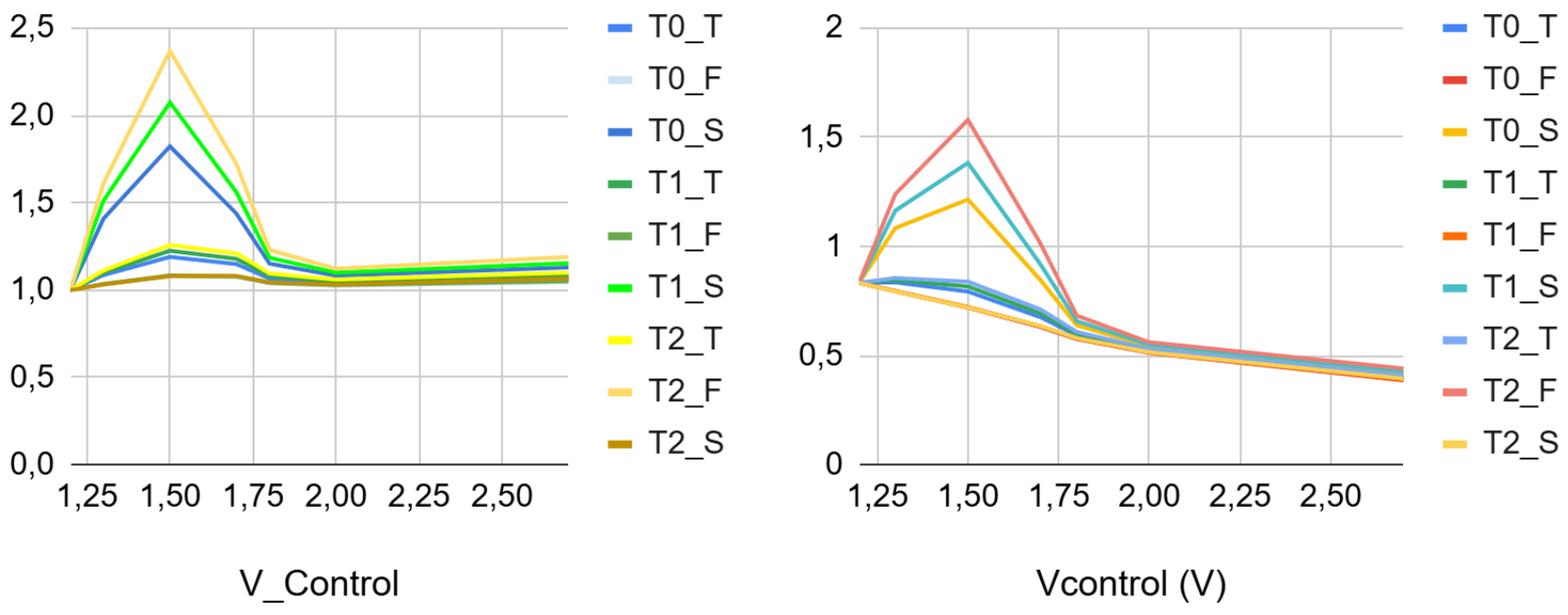}
\caption{Right: Normalized RMS signatures of the second harmonic across all PVT corners at $V_{DD}=3$~V. Normalization significantly reduces absolute variability while preserving sensitivity to control-voltage deviations. Left: Sensitivity of the signatures with respect to $V_{\text{CONTROL}}$ across all PVT corners.}
\label{fig:rms_norm}
\end{figure}

Fig.~\ref{fig:rms_combined} shows the absolute RMS values of the second harmonic across all PVT corners.
At $V_{\text{CONTROL}}=1.2$~V, values range from 0.04 (T2-F) to 0.22 (T0-F), a spread exceeding 5$\times$ between extreme corners.
However, the ordering with respect to $V_{\text{CONTROL}}$ is preserved for all corners with no curve crossings\blue{, except within the non-monotonic region identified in Section~\ref{sec:results}-B}---a critical property for fault detection.

The RMS signatures were normalized to the fault-free reference ($V_{\text{CONTROL}} = 1.2$~V at each corner).
The resulting relative signatures, shown in Fig.~\ref{fig:fault_injection}, illustrate the normalized RMS signatures for all nine PVT corners at the three representative operating points.
At $V_0=1.2$~V, signatures for slow-process corners (T0-S, T1-S, T2-F) clearly exceed $\tau_\text{high}$ for faults $|\Delta V| \geq 150$~mV, while fast-process corners remain within bounds for smaller faults.
At $V_0=2.0$~V, all corners remain within the [$\tau_\text{low}$, $\tau_\text{high}$] band across the entire fault range, confirming the unsuitability of this operating point.
The intermediate point $V_0=1.5$~V represents a trade-off, achieving meaningful FC for larger faults while remaining less sensitive to small parametric deviations.
These results confirm that biasing in the high-sensitivity region maximizes fault coverage and motivate the use of per-corner adaptive thresholds as a future improvement.

\subsubsection{Post-Calibration Temperature and Supply Drift}

\blue{Since calibration is performed once per chip at a single reference condition, post-calibration environmental drift---particularly temperature and supply voltage---can itself produce test escapes if it approaches or exceeds $\tau$.
To quantify this risk, the same physical device (fixed process corner) was evaluated across the three temperature points and three supply voltages with $V_\text{CONTROL}$ held fixed at $V_0$ and no fault applied; calibration was assumed at the T1 temperature corner and $V_{DD}=3.0$~V.}

\blue{Using $S_\text{HPF}$ alone, this drift is substantial and strongly process-dependent: averaged across all corners and operating points, temperature-induced drift ranges from 21.4\% to 67.0\%, and supply-induced drift averages 21.0--22.4\%, with the slow-process corner contributing the largest deviations (up to 275.9\% worst-case at $V_0=1.2$~V).
This finding motivates a refinement of the detection metric: instead of $S_\text{HPF}$ alone, the ratio
\begin{equation}
  R = \frac{S_\text{HPF}}{S_\text{LPF}}
\end{equation}
is used for the pass/fail decision in place of Eq.~(3)\footnote{Equation numbering refers to the methodology section; $R$ replaces $S_\text{norm}$ in Eq.~(4) of Section~\ref{sec:methodology}.}, since both filter outputs derive from the same VCO and are similarly affected by temperature and supply variations, allowing their ratio to substantially cancel common-mode environmental drift.}

\blue{Table~\ref{tab:temp_drift} reports the resulting drift using $R$.
Average temperature-induced drift drops to a consistent $\approx$10\% across all three operating points (from 21--67\% using $S_\text{HPF}$ alone), and average supply-induced drift drops to 0.4--2.3\% (from 21--22\%).
The combined worst-case drift, evaluated by varying temperature and supply simultaneously in their most adverse combination, remains dominated by the slow-process corner at high temperature, reaching 28--31\% across operating points regardless of which supply voltage produces the worst case; this indicates that supply variation does not introduce an independent failure mode beyond what is already captured by temperature drift once the ratio metric is used.}

\begin{table}[t]
\centering
\caption{Post-Calibration Drift (No Fault): $S_\text{HPF}$ Alone vs.\ Ratio Metric $R$, Average / Worst-Case}
\label{tab:temp_drift}
\begin{threeparttable}
\resizebox{\linewidth}{!}{%
\centering
\setlength{\arrayrulewidth}{0.4pt}
\begin{tblr}{
  colspec = {Q[45] Q[100] Q[85] Q[105] Q[85]},
  vline{2,3,4,5} = {2,3,4}{0.4pt},
  rowsep = 1.15pt,
  hline{1,2,5} = {1.2pt},
  hline{3,4} = {-}{0.4pt},
}
\textbf{$V_0$} &
\textbf{Temp., $S_\text{HPF}$} &
\textbf{Temp., $R$} &
\textbf{Supply, $S_\text{HPF}$} &
\textbf{Supply, $R$} \\
1.2~V & 67.0/275.9\% & 10.1/26.2\% & 22.4/32.7\% & 2.3/7.3\% \\
1.5~V & 29.5/95.8\%  & 10.6/29.8\% & 21.8/28.2\% & 0.4/0.9\% \\
2.0~V & 21.4/66.4\%  & 9.6/27.1\%  & 21.0/26.5\% & 1.1/1.5\% \\
\end{tblr}
}
\begin{tablenotes}\footnotesize
\item[] Entries indicate average/worst-case relative drift after calibration over the evaluated PVT space.
\end{tablenotes}
\end{threeparttable}
\end{table}

\blue{The residual drift under $R$ remains dominated by the slow-process corner (15--16\% on average, up to 30\% worst-case across combined temperature and supply variation), while typical-process devices remain within 8\% on average.
This indicates that the ratio metric provides reliable environmental robustness for typical-process devices, while slow-process devices may require additional compensation to fully eliminate the risk of test escapes.
On top of this circuit-level mitigation, many wearable systems already integrate a temperature sensor for biometric monitoring purposes; this existing sensor could be reused, at no additional hardware cost, to apply a coarse temperature-dependent correction or to flag operation outside the calibrated temperature range, further reducing residual risk for slow-process devices.
}

\subsection{Fault Sensitivity and Detection Capability}

Fig.~\ref{fig:rms_norm} shows the derivative of the normalized RMS signature with respect to $V_{\text{CONTROL}}$, indicating sensitivity to bias variations across all PVT corners.
Maximum sensitivity occurs in the lower control-voltage region (1.2--1.5~V): the normalized RMS increases from 1.0 at 1.2~V to 1.225 at 1.5~V in the T1-T corner, a 22.5\% change for a 300~mV shift.
At higher voltages (2.0--2.7~V) the change is only 3.2\% for a 700~mV shift.

In the high-sensitivity region ($V_{\text{CONTROL}} < 1.8$~V), fault-induced deviations clearly exceed the PVT envelope: a 300~mV shift produces $\approx$22\% normalized RMS change versus a $\leq$9.5\% PVT spread, exceeding $\tau_\text{high}$ with comfortable margin.
HPF outputs are more sensitive to localized waveform distortions, while LPF outputs provide a stable reference; their combined use improves fault observability without increasing circuit complexity.
At $V_0=1.2$~V, FC reaches 65\% for $|\Delta V|\geq 150$~mV, rising to 69\% and 100\% for 300~mV and 450~mV respectively, confirming the practical detection capability in the high-sensitivity region\blue{, with the caveat on negative-polarity coverage noted in Section~\ref{sec:results}-B}.

\subsection{Comparison with ADC-Based Voltage Monitoring}

\begin{table}[t]
\centering
\caption{Comparison with Flexible ADCs}
\label{tab:adc_comp}
\scalebox{1}{\input{tables/comparison}}
\end{table}

Table~\ref{tab:adc_comp} compares the proposed test circuit against representative ADC-based monitoring solutions in the same IGZO-based FE technology.
The Flash and Binary Search ADCs were designed and post-layout simulated at 3~V; the Binary Search ADC is derived from the fault-tolerant design in~\cite{lozano2026faulttolerantdesignigzobased} and resized for 3~V operation, while~\cite{Alkhalil:BioCAS:2022:FlexibleSAR} already targets 3~V.
\blue{A second-order $\Sigma\Delta$ ADC implementation in IGZO TFTs has also been reported~\cite{correia2022sigmadelta}, achieving 65~dB dynamic range at $V_{DD}=10$~V, but requiring 22~mW and 10~mm$^2$---several orders of magnitude larger than the proposed circuit.
While the higher supply voltage partially explains this overhead, it also reflects the inherent complexity of oversampled architectures, which require additional clock generation and digital decimation filtering unsuitable for low-frequency, low-power FE systems.}
The proposed circuit achieves the lowest power (0.017~mW) and the smallest area (0.0064~mm$^2$) across all compared solutions, representing a 26$\times$ area reduction over the most compact ADC and nearly 1500$\times$ over the SAR implementation.
\blue{The comparison in Table~\ref{tab:adc_comp} is symmetric in scope: the reported area for both the proposed circuit and the ADC references excludes downstream post-processing---RMS extraction and normalization in this work, and threshold comparison logic in the ADC case---.}
This comparison nonetheless understates the true ADC overhead, since ADC accuracy is itself highly sensitive to supply-voltage variations, comparator matching, and reference stability, whereas the proposed signature remains robust to these effects by construction, as quantified in Section~\ref{sec:results}-E.

\subsection{Test Circuit Overhead Analysis}

\begin{table}[t]
\centering
\caption{Comparison with representative FE AMS systems}
\label{tab:fe_systems}
\resizebox{\linewidth}{!}{%
\begin{tblr}{
  colspec = {Q[35] Q[85] Q[80] Q[65]},
  rowsep = 1.15pt,
  hline{1,2,5} = {1.2pt},
  hline{3,4} = {-}{0.4pt},
}
\textbf{Ref} & 
\textbf{Area (mm$^2$)} &
\textbf{Power (mW)} &
\textbf{Overhead} \\
\cite{FlexRISCV:MICRO25}    & 0.59--1.02  & 0.71--0.91  & $<$1.1\% \\
\cite{FlexCoProc:DATE26} & 2.41--2.44 & 1.49--1.53 & $<$0.3\% \\
\cite{FlexClassifier:ICCAD25}    & 0.20--27.6  & 0.14--20.3  & $<$3.2\% \\
\end{tblr}
}
\vspace{-2ex}
\end{table}

Table~\ref{tab:fe_systems} lists representative IGZO AMS target systems.
The proposed circuit occupies 0.0064~mm$^2$, representing an overhead below 3.2\% relative to the smallest listed system and below 0.07\% relative to the largest, confirming that the test circuit overhead is negligible with respect to realistic FE CUTs.

The proposed frequency-domain approach detects not only DC-bias errors but also changes in AC loading, bandwidth, or slew rate that are invisible to DC-voltage measurements.
\textit{For applications where a VCO is already present, the incremental test circuit overhead is minimal, making this approach particularly attractive for resource-constrained FE systems.}

%% file: tables/sizing.tex
\resizebox{\linewidth}{!}{%
\centering
\setlength{\arrayrulewidth}{0.4pt} 
\begin{tblr}{
  colspec = {Q[25] Q[25] Q[25]}, 
  vline{2,3} = {-}{0.4pt},
  hline{1,2,12} = {1.2pt}, 
  rowsep = 1.1pt, 
  hline{8,10} = {-}{0.4pt}, 
  hline{3,4,5,6,7,9,10,11} = {2,3}{0.4pt}, 
}
\textbf{Component} & \textbf{Element} & \SetCell[c=1]{c} \textbf{Size} \\ 

\textbf{VCO} & R0\&R10 & r = 50$k\Omega$\\
 & R1:R7 & r = 172.47$k\Omega$\\
 & R8 & r = 120$k\Omega$\\
 & R9 & r = 5.6$M\Omega$\\
 & M0:M4 & W= 20$\mu$m L= 600nm \\
 & C & cap = 0.3pF \\
\textbf{HPF} & R & r = 251.27$k\Omega$\\
 & C & cap = 2pF\\
\textbf{LPF} & R & r = 172.47$k\Omega$\\
 & C & cap = 2pF\\

\end{tblr}
\vspace{-3ex}
}

%% file: tables/comparison.tex
\begin{threeparttable}
\resizebox{\linewidth}{!}{%
\centering
\setlength{\arrayrulewidth}{0.4pt} 

\begin{tblr}{
  colspec = {Q[105] Q[70] Q[70] Q[70]}, 
  vline{2} = {2,3,4,5,6}{0.4pt},
  rowsep = 1.15pt, 
  hline{1,2,7} = {1.2pt},
  hline{3,4,5,6} = {-}{0.4pt},
}
\textbf{Ref} & \textbf{Resolution} & \textbf{Power (mW)} & \textbf{Area(mm$^2$)} \\
This work & - & 0.017 & 0.0064 \\
Binary ADC~\cite{lozano2026faulttolerantdesignigzobased} & 3 & 0.071 & 0.17 \\
Flash ADC & 4 & 0.126 & 1.18 \\
$\Sigma\Delta$ ADC~\cite{correia2022sigmadelta} & - & 22\tnote{*} & 10\tnote{*} \\
SAR ADC~\cite{Alkhalil:BioCAS:2022:FlexibleSAR} & 7.8 & 2.93 & 9.4 \\
\end{tblr}
}
\begin{tablenotes}\footnotesize
    \item[]$*$ Measured with $V_{dd}$ 10V.
\end{tablenotes}
\end{threeparttable}

%% file: section/5_Conclusion.tex
\section{Conclusion}
\label{sec:conclusion}

Unipolar TFT technologies face significant testability challenges due to device variability, post-manufacturing susceptibility to mechanical failures, and the cost and in-field inaccessibility of specialized ATE.
This work presents a frequency-domain test signature generation circuit exploiting spectral energy distribution as a lightweight test observable for AMS CUTs in flexible electronics.
A VCO converts control-voltage deviations into frequency shifts processed through analog filters, with RMS values of harmonic components serving as compact signatures exhibiting monotonic sensitivity across PVT corners.
Fault modeling across nine PVT corners and three operating points demonstrates that in the high-sensitivity region ($V_0=1.2$~V), FC reaches 65\% for $|\Delta V|\geq 150$~mV, 69\% for $|\Delta V|\geq 300$~mV, and 100\% for $|\Delta V|=450$~mV under a two-sided threshold ($\tau_\text{high}=1.10$, $\tau_\text{low}=0.90$)\blue{, with negative-polarity coverage at $V_0=1.2$~V subject to a verified non-monotonic region below $V_\text{CONTROL}=1.125$~V}.
The circuit occupies 6390~$\mu$m$^2$ and consumes 16.7~$\mu$W, achieving 26$\times$ area reduction versus compact ADCs and nearly 1500$\times$ versus SAR implementations, with area overhead below 3.2\% relative to representative FE target systems.
The generated signature is readable via two standard I/O pins by a microcontroller present in the wearable system, enabling in-field test without ATE.
Though validated using IGZO TFTs on PragmatIC's FlexIC process, the methodology applies to any n-type-only technology.